\documentclass[aps,prl,twocolumn,showpacs,superscriptaddress]{revtex4}
\usepackage{graphicx}
\begin{document}


\title{Static magnetic order of Sr$_{4}$A$_{2}$O$_{6}$Fe$_{2}$As$_{2}$ 
(A = Sc and V) revealed by local probes\/}
     \author{J. Munevar}
     \affiliation{Centro Brasileiro de Pesquisas Fisicas, Rua Xavier Sigaud 150, Rio de Janeiro, Brazil}
     \author{D.~R.~S\'anchez}
     \affiliation{Instituto de F\'isica, Universidade Federal Fluminense, 
24210-346 Niter\'oi, RJ, Brazil}
     \affiliation{Centro Brasileiro de Pesquisas Fisicas, Rua Xavier Sigaud 150, Rio de Janeiro, Brazil}
     \author{M. Alzamora}
     \author{E. Baggio-Saitovitch}
     \affiliation{Centro Brasileiro de Pesquisas Fisicas, Rua Xavier Sigaud 150, Rio de Janeiro, Brazil}
     \author{J.~P.~Carlo}
     \affiliation{Department of Physics, Columbia University, New York, New York 10027, USA}
     \author{T.~Goko}
     \affiliation{Department of Physics, Columbia University, New York, New York 10027, USA}
     \affiliation{TRIUMF, 4004 Wesbrook Mall, Vancouver, B.C., V6T 2A3, Canada}
     \author{A.~A.~Aczel}
     \author{T.~J.~Williams}
     \affiliation{Department of Physics and Astronomy, McMaster University, Hamilton, Ontario L8S 4M1, Canada}
     \author{G.~M.~Luke}
     \affiliation{Department of Physics and Astronomy, McMaster University, Hamilton, Ontario L8S 4M1, Canada}
     \affiliation{Canadian Institute of Advanced Research, Toronto, Ontario M5G 1Z8, Canada}
     \author{Hai-Hu Wen}
     \author{Xiyu Zhu}
     \author{Fei Han}
     \affiliation{National Laboratory for Superconductivity, Institute of Physics and Beijing National Laboratory for Condensed Matter Physics, Chinese Academy of Sciences, P.O. Box 603, Beijing 100190, 
People's Republic of China}
     \author{Y.~J.~Uemura}
     \altaffiliation[author to whom correspondences should be addressed: E-mail
tomo@lorentz.phys.columbia.edu]{}
     \affiliation{Department of Physics, Columbia University, New York, New York 10027, USA}
     \date{\today}
   \begin{abstract}
\noindent
{Static magnetic order of quasi two-dimensional FeAs compounds
Sr$_{4}$A$_{2}$O$_{6-x}$Fe$_{2}$As$_{2}$, with A = Sc and V, has been
detected by $^{57}$Fe M\"ossbauer and muon spin relaxation ($\mu$SR) spectroscopies.
The non-superconducting stoichiometric ($x=0$) A = Sc system exhibits a static internal/hyperfine magnetic field
both at the $^{57}$Fe and $\mu^{+}$ sites, indicating antiferromagnetic order of Fe moments
below $T_{N}$ = 35 K with $\sim 0.1$ Bohr magneton per Fe at T = 2 K. The superconducting and oxygen deficient
($x=0.4$) A = V system exhibits a static internal field only at the $\mu^{+}$ site below
$T_{N} \sim 40$ K, indicating static magnetic order of V moments co-existing with superconductivity without
freezing of Fe moments.
These results suggest that the 42622 FeAs systems belong to the same paradigm with the 1111 and 122 FeAs
systems with respect to magnetic behavior of Fe moments.\/}
\end{abstract}
\pacs{
74.70.Xa 
75.30.-m 
76.80.+y 
76.75.+i 
}
\maketitle

Since the discovery in the spring of 2008 \cite{hosono}, iron arsenide superconductors and related systems
have generated a burst of research. 
These compounds have a FeAs layer together with a charge reservoir layer composed of  
RO (with R as a rare earth) in the ``1111' systems 
\cite{hosono,oldrefs26,daineutron,daiphase,musrphase}, alkali
atoms (Ca, Sr, Ba) in the ``122'' materials \cite{122a,122b,122c}, and Li and Na atoms 
in the ``111'' compounds \cite{jin111,blundell111}.  In the 
1111, 122, and Na-based 111 FeAs compounds, 
parent systems exhibit antiferromagnetic order, and superconductivity
arises through carrier doping via chemical substitutions and/or 
application of external pressure. 
Recently, several groups reported synthesis of a new group of FeAs systems
separated by perovskite layers such as Sr$_{2}$AO$_{3}$, leading to 
the ``42622'' systems Sr$_{4}$A$_{2}$O$_{6-x}$Fe$_{2}$As$_{2}$ 
\cite{0903.5124,0904.1732,0910.1537,0911.0450}, 
where A denotes Sc, Ti, Cr, V, and other transition metal
atoms.  Subsequently discovered were ``32522'' 
Sr$_{3}$Sc$_{2}$O$_{5}$Fe$_{2}$As$_{2}$ \cite{0905.0337}, 
the homologous series of 
Ca$_{n+1}$M$_{n}$O$_{y}$(Fe$_{2}$As$_{2}$) [n = 3, 4, 5,
$y \sim$ 3n-1, M = (Sc,Ti) and (Mg,Ti)] \cite{1009.5491},
and (Ca$_{n+2}$(Al,Ti)$_{n}$O$_{y}$)(Fe$_{2}$As$_{2}$) [n = 2,3,4] 
\cite{1008.2582}.
Despite a large distance between FeAs layers separated by the perovskite
layers, many of these ``perovskite + FeAs'' (perov-FeAs) systems exhibit superconductivity, with 
the $T_{c}$ reaching as high as 47 K \cite{1009.5491}.   

Study of magnetic order of perov-FeAs systems, however, has been difficult, because:
(1) non-superconducting compounds do not exhibit clear signature of the spin density wave 
(SDW) ordering of Fe in resistivity nor crystal structure, most-likely due to weak magnetic coupling between 
FeAs layers separated by a larger interplane distance;  
(2) it is often difficult to separate magnetism of Fe from that of V, Cr, and other
transition-metal elements in these systems; and (3) neutron scattering and magnetization
measurements do not provide information on the volume fraction of magnetically ordered regions.
No signature of static order of Fe was detected by the $^{57}$Fe M\"ossbauer effect in 
Ba$_{4}$Sc$_{2}$O$_{6}$Fe$_{2}$As$_{2}$ \cite{0904.0479} and 
Sr$_{3}$Sc$_{2}$O$_{5}$Fe$_{2}$As$_{2}$ \cite{0905.0337}, and magnetic
order of Fe has so far not been reported in any of the perov-FeAs systems to our knowledge.
Static magnetic order of the transition metal atom A in Sr$_{4}$A$_{2}$O$_{6}$Fe$_{2}$As$_{2}$
(A = Cr, V) was reported from neutron 
scattering \cite{0911.0450,1008.2687} and the $^{57}$Fe M\"ossbauer effect 
\cite{0904.0479} for the A = Cr compound, 
and from X-ray absorption and M\"ossbauer \cite{1007.3980} measurements for the A = V compound.
The results on the A = V compound, however, are not conclusive due to limited neutron signal quality
on polycrystalline specimens in ref. \cite{1008.2687}, 
and limited temperature range (T $>$ 20 K), 
indirect signature relying on the absence of Fe ordering, 
and lack of volume-related information
in ref. \cite{1007.3980}. 

In this paper, we present a combination of $^{57}$Fe M\"ossbauer effect and positive 
muon spin relaxation ($\mu^{+}$SR) measurements, which clearly reveal that:
(a) non-superconducting Sr$_{4}$Sc$_{2}$O$_{6}$Fe$_{2}$As$_{2}$ exhibits antiferromagnetic
SDW order of Fe moments; (b) 
superconducting Sr$_{4}$V$_{2}$O$_{5.6}$Fe$_{2}$As$_{2}$ exhibits static
magnetic order of V moments (without ordering of Fe sub-lattice), coexisting with superconductivity;
and (c) magnetic order in both of these systems develops in the full volume fraction.

Polycrystalline sintered specimens of stoichiometric Sr$_{4}$Sc$_{2}$O$_{6}$Fe$_{2}$As$_{2}$
and oxygen deficient Sr$_{4}$V$_{2}$O$_{5.6}$Fe$_{2}$As$_{2}$ (samples \# 1 - \# 3) 
were synthesized in Beijing
following the methods described in refs. \cite{0904.1732,0910.1537}.
The A = V sample was checked by magnetization measurements,
which confirmed superconductivity below T$\sim$ 25 K (Figs. 3(c) and 4(e))
and hysteresis corresponding to a very weak ferromagnetic polarization
(Fig. 2(c)) at T = 5 K.  $^{57}$Fe M\"ossbauer effect and $\mu$SR measruements
were carried out, respectively, at CBPF (Rio, Brazil) and 
TRIUMF (Vancouver, Canada) using the same specimens.

Figure 1 compares M\"ossbauer spectra obtained in the A = Sc and
A = V compounds.  At room temperature, the A = Sc system shows
a unique doublet with the quadrupole splitting value
$\Delta E_Q=-0.165(5)$, indicating a single paramagnetic site.  
As the sample is cooled down below 40 K the doublet 
splits in two phases, a remaining paramagnetic doublet and a magnetic sextet, 
indicated by the light-blue and dark-blue lines, respectively, 
in Fig. 1(a).   At T = 4.3 K, the paramagnetic signal disappears and the 
hyperfine field for the magnetic site becomes 1.73(2) T, 
which corresponds to a static Fe moment of 0.12(1) $\mu_B$.  The angle $\theta$ of the moment
direction fits to $\sim$ 80 degrees, 
indicating that the Fe moments align almost perpendicular to the crystallographic $c$ axis.  
The isomer shift values $\delta=0.465 (3)$ mm/s at T = 300 K and $0.426(4)$ mm/s at T = 4.3 K 
indicate that the valence state and electronic configuration surrounding an Fe atom are nearly independent of 
temperature. 
Figures 2(a) and (b) show the hyperfine field and the volume fraction of
the magnetically ordered region, derived by fitting the sextet signal in the A = Sc compound.

\begin{figure}[t]

\begin{center}
\includegraphics[angle=0,width=3.0in]{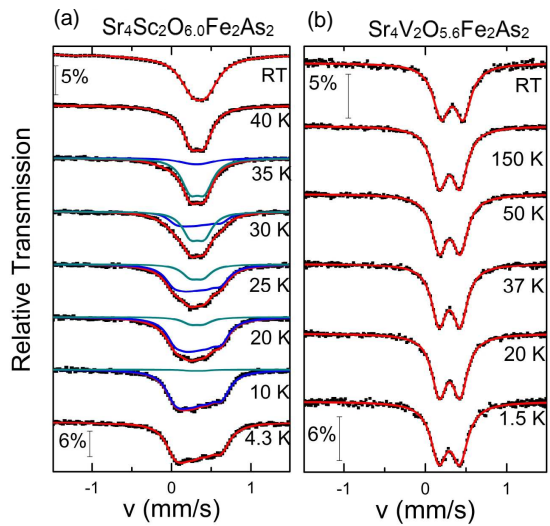} 
\label{Figure 1.} 
\caption{\label{Figure 1.} 
$^{57}$Fe M\"ossbauer effect spectra in 
(a) non-superconducting Sr$_{4}$Sc$_{2}$O$_{6}$Fe$_{2}$As$_{2}$, and (b)
superconducting Sr$_{4}$V$_{2}$O$_{5.6}$Fe$_{2}$As$_{2}$ (\# 3).
The spectra in (a) have been fitted as a sum of the doublet
signal from the paramagnetic Fe (light blue line) and the sextet signal from
the ordered static Fe moments (dark blue line).  The line width in (b) is nearly
independent of temperature for T = 20 - 300 K.}
\end{center}
\end{figure}

Figure 1(b) shows
the M\"ossbauer spectra for Sr$_4$V$_2$O$_{5.6}$Fe$_2$As$_2$, from room temperature down to 1.5 K.  
The well-defined absorption doublet has an isomer shift $\delta=0.411(2)$ mm/s and a quadrupole splitting 
$\Delta E_Q=-0.256(2)$ mm/s typical for low spin Fe$^{+2}$.  These values are nearly independent 
of temperature, suggesting the absence of structural phase transition, in agreement with previous structural studies
\cite{0904.1732,0910.1537}.  The linewidth exhibits almost no change 
with decreasing temperature, demonstrating that Fe moments
do not participate in static magnetic order in the A = V 42622 perov-FeAs compound.  To estimate
the static hyperfine field in this system, we subtracted the linewidth at room temperature from the width observed
at low temperatures, and plot the result in Fig. 2(a).  This figure clearly demonstrates that the A = Sc compound exhibits 
static order of Fe moments, but the A = V system does not.  
The linewidth $\Gamma$ shows a small increase at very lower temperatures: 
$\Gamma=0.223(4)$ mm/s at T = 4.3 K while $\Gamma=0.238(7)$ mm/s at T = 1.5 K.  This could be due to an 
effective hyperfine field generated by the ordering of V moments.
As shown in Fig. 2(c), the A = V system exhibits hysteresis in the magnetization.  This indicates a small
ferromagnetic component associated with static ordering of V moments.  The observed 
spontaneous ferromagnetic polarization $\sim$ 0.01 emu/g
at H = 0 for T = 5 K, however, corresponds only to $7\times 10^{-4}$ Bohr magneton per V atom. 

\begin{figure}[t]

\begin{center}
\includegraphics[angle=0,width=3.20in]{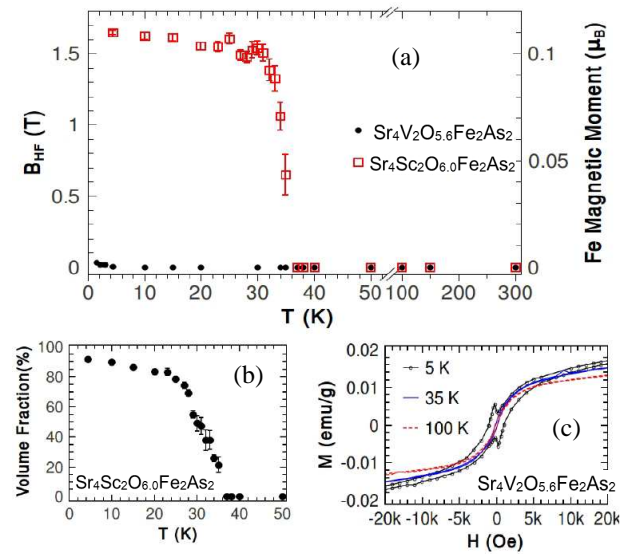}
\label{Figure 2.} 
\caption{\label{Figure 2.} 
(a) Static hyperfine field and corresponding Fe moment size in
non-superconducting Sr$_{4}$Sc$_{2}$O$_{6}$Fe$_{2}$As$_{2}$, and
superconducting Sr$_{4}$V$_{2}$O$_{5.6}$Fe$_{2}$As$_{2}$ (\# 3), estimated from the linewidth
in $^{57}$Fe M\"ossbauer effect.  (b) The volume fraction of ordered Fe in 
Sr$_{4}$Sc$_{2}$O$_{6}$Fe$_{2}$As$_{2}$ derived by fitting the spectra as a sum of
the doublet line representing paramagnetic Fe and the sextet line for ordered Fe moments.
(c) Magnetization hysteresis observed in Sr$_{4}$V$_{2}$O$_{5.6}$Fe$_{2}$As$_{2}$
(\# 3).}
\end{center}
\end{figure}

Figure 3 compares $\mu$SR time spectra observed in zero field in the A = Sc and V compounds.  In (a)
the Sc compound exhibits an onset of precession below T $\sim$ 30 K, which is a direct indication of
static magnetic order.  The temperature dependence of the precession frequency $\nu(T)$, shown in 
Fig. 4(a), is consistent with the results of the M\"ossbauer hyperfine field in Fig. 2(a).  
In parent antiferromagnetic compounds of the 122 and 1111 FeAs systems, there exists a nearly 
linear relationship between the ordering temperature $T_{N}$ and the 
ZF-$\mu$SR frequency $\nu (T\rightarrow 0)$ at low temperatures \cite{uemurasces},
which is proportional to the magnitude of ordered Fe moments.   As shown in Fig. 4(b),
the results for the A = Sc 42622 system clearly follow 
this trend, demonstrating commonalities of Fe magnetism among parent 
compounds of all of these FeAs systems.  The continuous variation of the 
ordered Fe moment size suggests that this magnetism is likely arising from 
itinerant Fe electrons, though in some cases localized spin systems exhibit similar behavior
due to low dimensionality and/or frustration \cite{kojimaprl}.  
The ordered Fe moment size $\sim$ 0.1 Bohr magneton, 
estimated by $\mu$SR from relative magnitudes of $\nu(T\rightarrow 0)$, 
is consistent with the M\"ossbauer results.  Figure 4(c) shows 
the volume fraction of the ordered region estimated by the amplitude of the oscillating $\mu$SR signal.  
This figure, together with the M\"ossbauer results in Fig. 2(b),
indicate a gradual build up of static magnetic order of Fe moments below $T_{N}\sim 35$ K 
in the non-superconducting A = Sc system. 

\begin{figure}[t]

\begin{center}
\includegraphics[angle=0,width=3.25in]{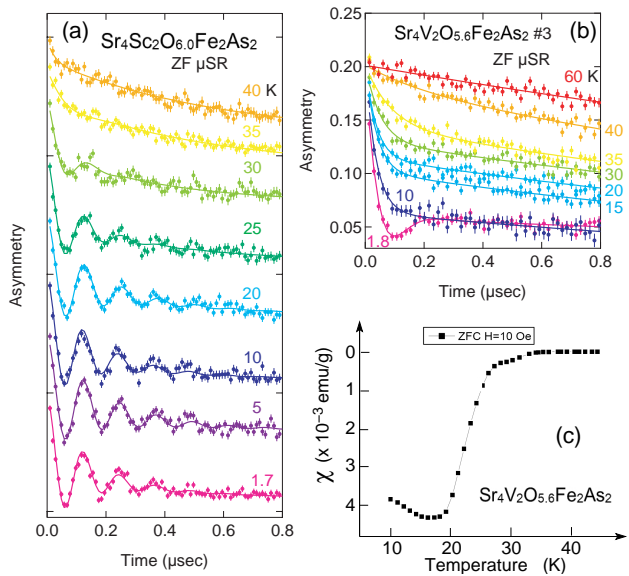}
\label{Figure 3.} 
\caption{\label{Figure 3.} 
Time spectra of zero-field $\mu$SR measurements in 
(a) Sr$_{4}$Sc$_{2}$O$_{6}$Fe$_{2}$As$_{2}$ and (b) Sr$_{4}$V$_{2}$O$_{5.6}$Fe$_{2}$As$_{2}$ 
(\#3).  
Separate measurements in longitudinal field confirmed that the observed
oscillation and relaxation are due to static magnetic fields.  
(c) shows magnetization results on Sr$_{4}$V$_{2}$O$_{5.6}$Fe$_{2}$As$_{2}$ (\#3),
which confirmed the onset of superconductivity at $T \sim 25$ K.}
\end{center}
\end{figure}

\begin{figure}[t]
\begin{center}
\includegraphics[angle=0,width=3.25in]{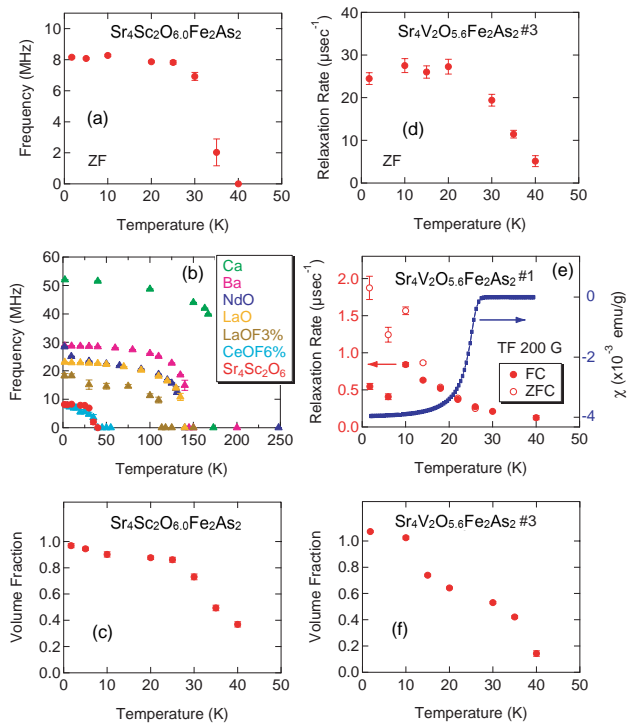}
\label{Figure 4.} 
\caption{\label{Figure 4.}
(a) Muon spin precession frequency in zero field observed in
Sr$_{4}$Sc$_{2}$O$_{6}$Fe$_{2}$As$_{2}$.
(b) Comparison of this frequency with those observed in other
FeAs systems \cite{uemurasces}.
(c) Volume fraction of regions with static magnetic order
derived from the oscillating amplitude of the $\mu$SR signal in 
Sr$_{4}$Sc$_{2}$O$_{6}$Fe$_{2}$As$_{2}$.
(d) Muon spin relaxation rate of the fast-relaxing component 
in zero field due to static magnetic order, (e) the relaxation rate of the
slow-relaxing component from the paramagnetic volume observed in A transverse
field of 200 G, and (f) the volume fraction of the magnetically ordered region, 
in superconducting Sr$_{4}$V$_{2}$O$_{5.6}$Fe$_{2}$As$_{2}$.
The results in (d) and (f) were obtained on sample \# 3, while in (e) on 
sample \# 1: both having nominally identical compositions.  
Magnetic susceptibility of sample \# 1 in (e) shows the onset of 
superconductivity.} 
\end{center}
\end{figure}

We also performed $\mu$SR studies on oxygen deficient specimens of the A = V compound.  
Susceptibility results in Fig. 3(c) confirm superconductivity of our specimen \# 3 with 
a decent Meissner signal.  The zero-field $\mu$SR time spectra in Fig. 3(b) exhibit the
onset of a fast relaxation, causing damping of the asymmetry within t = 200 ns, below 
T $\sim$ 40 K. With a separate $\mu$SR study in a longitudinal field, we confirmed 
that this relaxation is due to static random fields.  Figures 4(e) and (g) 
show the relaxation rate of the fast-relaxing signal and the volume fraction of the magnetically ordered
region derived from the amplitude of that signal.  By comparing Fig. 3(a) for the A = Sc system
and Fig. 3(b) for the V system, we notice that the rate of initial damping of the 
asymmetry is comparable.  This indicates that the static internal fields
at the muon site in these two systems are comparable in magnitude.  Although
it is not possible to provide a more precise estimate
due to lack of information on the location of the muon site and the exact spin configuration 
of the V moments, 
this observation points towards $\sim$ 0.1 Bohr magneton as an order of magnitude estimate for 
the static V moment at $T \rightarrow 0$.  This finding, together with the small magnetic hysteresis loop measured, 
suggest dominantly 
antiferromagnetic spin correlations of V moments, consistent with recent neutron 
measurements \cite{1008.2687}.  
We note that commensurate antiferromagnetic order was observed by neutron scattering 
in a stoichiometric A = Cr 42622 perov-FeAs compound with an ordered 
moment size of 2.75 Bohr magneton per Cr \cite{0911.0450}.  

In the A = V compound, between T = 15 - 40 K, a significant volume fraction remains without 
having a large internal field from static V moments.  In separate $\mu$SR measurements in a
transverse external field (TF) of 200 G, we attempted to measure the magnetic 
field penetration depth from the signal
representing such ``paramagnetic/nonmagnetic'' volume fraction.  As shown in Fig. 4(e), 
we found different TF $\mu$SR
relaxation rates between the field-cooling (FC) and zero-field-cooling (ZFC) procedures, which is a 
typical response for superconductors relevant to pinning of flux vortices \cite{lpleprlbedt}.
The scattering of data points below T = 15 K in Fig. 4(e) is due to a decreasing 
``paramagnetic'' volume and difficulty in separating the relaxation due to magnetic order
from the effect of flux vortices.  The FC relaxation rate $\sigma \sim 1 \mu$s$^{-1}$ at $T\rightarrow 0$
is comparable to the rate observed in superconducting 1111 and 122 FeAs systems \cite{uemurasces}.
These results indicate that superconductivity survives at least in the volume without 
static order of V.  Although our $\mu$SR results do not give direct information on 
whether or not superconductivity and magnetic order co-exist in the same volume,
the decent Meissner signal and essentially full volume fraction of the  
magnetically ordered region at $T \rightarrow 0$ suggest that static magnetic
order in V sublattices may not give any adverse effect on superconductivity in the 
FeAs layers.   

In summary, the combined M\"ossbauer and $\mu$SR measurements demonstrated 
static magnetic order of the Fe sub-lattice in the A = Sc compound and the V sub-lattice in 
the A = V 42622 FeAs compound.
Contrary to some theoretical proposals \cite{mazin,pickett} suggesting that the
42622 systems may be very different from other FeAs systems with respect
to magnetism of the Fe sub-lattices and fermi-surface
nesting, the present results indicate that the 42622 systems
belong to the same paradigm with other FeAs systems, exhibiting static
antiferromagnetic order of Fe moments in the parent non-superconducting compound
and superconductivity without static order of the Fe sub-lattice in a carrier-doped superconducting
compound. Magnetic order of the
transition-metal sub-lattice (V or Cr) in the perovskite layer does not seem to alter essential 
features of the FeAs layers, similarly to the case of the ordering of the rare-earth moments
in the 1111 systems.  The quasi two-dimensional feature of the 42622 systems seems to 
work against static antiferromagnetic order of the Fe moments.  This suppression of 
competing magnetic order might provide an indirect effect for promoting superconductivity 
in various perov-FeAs systems and lead to rather high $T_{c}$'s
of newly discovered homologous systems which have FeAs interlayer distances
even larger than those of the 42622 systems.

{\bf Acknowledgement:\/}  This work has been supported by the
US NSF under the Materials World Network (MWN: DMR-0502706 and 0806846) and
the Partnership for International Research and education (PIRE: OISE-0968226) programs at 
Columbia, by Canadian NSERC and CIFAR at McMaster,
and by CNPq and Faperj at CBPF in Rio, Brazil and NSFC and MOST of
China: 973 project 2011CB605900 at IOP in Beijing.\\

\vfill \eject
\end{document}